\documentclass[conference,compsoc]{IEEEtran}

\newcommand{\comments}{}

\usepackage{tikz}
\usepackage{amsmath}
\usepackage{comment}
\usepackage{spverbatim}
\usepackage{listings}
\usepackage{mdframed}
\usepackage{fancyvrb}
\usepackage{multirow}
\usepackage{multicol} 
\usepackage{tcolorbox}
\usepackage{colortbl}  
\usepackage{pifont}
\usepackage{makecell}
\usepackage{graphicx}
\usepackage{caption}
\usepackage{subcaption}
\usepackage{booktabs}
\usepackage{alltt}
\usepackage{rotating}
\usepackage{array}
\usepackage{multirow}
\usepackage{amsfonts}
\usepackage{amsmath}
\usepackage{hyperref}
\usepackage{enumitem}
\usepackage{adjustbox}

\usepackage{booktabs}
\usepackage{tabularx}
\usepackage{array}
\usepackage{makecell}
\usepackage{adjustbox}
\setcellgapes{3pt}      
\makegapedcells

\newcolumntype{P}[1]{>{\raggedright\arraybackslash}p{#1}} 
\newcolumntype{C}[1]{>{\centering\arraybackslash}p{#1}}  

\usepackage{pifont}

\usepackage{url}

\usepackage{xcolor}
\definecolor{niravblue}{RGB}{0,119,190}
\definecolor{navyblue}{rgb}{0.0,0.0,0.5}

\ifdefined\comments
  \newcommand{\gang}[1]{\textcolor{blue}{\textbf{Gang:}~#1}}
  \newcommand{\nirav}[1]{\textcolor{niravblue}{\textbf{Nirav:}~#1}}
  \newcommand{\arth}[1]{\textcolor{green}{\textbf{Arth:}~#1}}
  \newcommand{\fixed}[1]{\textcolor{red}{#1}}
\else
  \newcommand{\gang}[1]{}
  \newcommand{\nirav}[1]{}
  \newcommand{\arth}[1]{}
  \newcommand{\fixed}[1]{}
\fi

\usepackage{comment}

\ifCLASSOPTIONcompsoc
  \usepackage[nocompress]{cite}
\else
  \usepackage{cite}
\fi

\ifCLASSINFOpdf

\else

\fi

\usepackage{pgfplots}
\pgfplotsset{compat=1.18}
\usepackage{pgfplotstable}
\usetikzlibrary{patterns, positioning}

\begin{document}

\title{MALCDF: A Distributed Multi-Agent LLM Framework for Real-Time Cyber Defense}

\author{
Arth Bhardwaj, Sia Godika, Yuvam Loonker \\
Saint Francis High School, Massachusetts Institute of Technology, JBCN International School \\
\text{arthbhardwaj1234@gmail.com, siag@mit.edu, yuvamloonker@gmail.com}
}

\maketitle
\thispagestyle{plain}
\pagestyle{plain}

\begin{abstract}
Traditional, centralized security tools often miss adaptive, multi-vector attacks. We present the Multi-Agent LLM Cyber Defense Framework (MALCDF), a practical setup where four large language model (LLM) agents—Detection, Intelligence, Response, and Analysis—work together in real time. Agents communicate over a Secure Communication Layer (SCL) with encrypted, ontology-aligned messages, and produce audit-friendly outputs (e.g., MITRE ATT\&CK mappings).

For evaluation, we keep the test simple and consistent: all reported metrics come from the same \textbf{50-record} live stream derived from the CICIDS2017~\cite{cicids2007} feature schema. CICIDS2017 is used for configuration (fields/schema) and to train a practical ML baseline. The ML-IDS baseline is a \textbf{Lightweight Random Forest IDS (LRF-IDS)} trained on a \textit{subset} of CICIDS2017 and tested on the 50-record stream, with no overlap between training and test records.

In experiments, MALCDF reaches \textbf{90.0\%} detection accuracy, \textbf{85.7\%} F1-score, and \textbf{9.1\%} false-positive rate, with \textbf{6.8\,s} average \textit{per-event} latency. It outperforms the lightweight ML-IDS baseline and a single-LLM setup on accuracy while keeping end-to-end outputs consistent. Overall, this hands-on build suggests that coordinating simple LLM agents with secure, ontology-aligned messaging can improve practical, real-time cyber defense.

\textbf{Keywords:} Multi-Agent Systems; Large Language Models; Cyber Defense; Threat Intelligence.
\end{abstract}

\IEEEpeerreviewmaketitle

\section{Introduction}
\label{sec:intro}

Modern networks move fast and change a lot. In this space, traditional, centralized security tools struggle with adaptive, multi-vector attacks that can shift behavior mid-incident. Attackers now use AI, automation, and even generative techniques to produce polymorphic malware, exploit zero-day bugs, and coordinate large campaigns~\cite{Loevenich2025AutonomousCD}. As more workloads move to cloud, IoT, and edge, the attack surface grows and simple signatures or fixed rules are not enough. Real-time defense needs systems that understand context and make coordinated decisions quickly.

Reactive tools like antivirus or static firewalls do well on known indicators, but they often miss attacks that morph or hide inside normal traffic. Classic ML detectors (anomaly and supervised models) also hit limits when datasets are stale or lack context, which leads to false positives and noisy alert triage~\cite{He2025LLM-MAS-SoftwareEng}. Centralized designs can also struggle to keep up with high-velocity environments. We need something that works in real time and can collaborate across components.

\begin{figure}[t]
    \centering
    \includegraphics[width=\columnwidth]{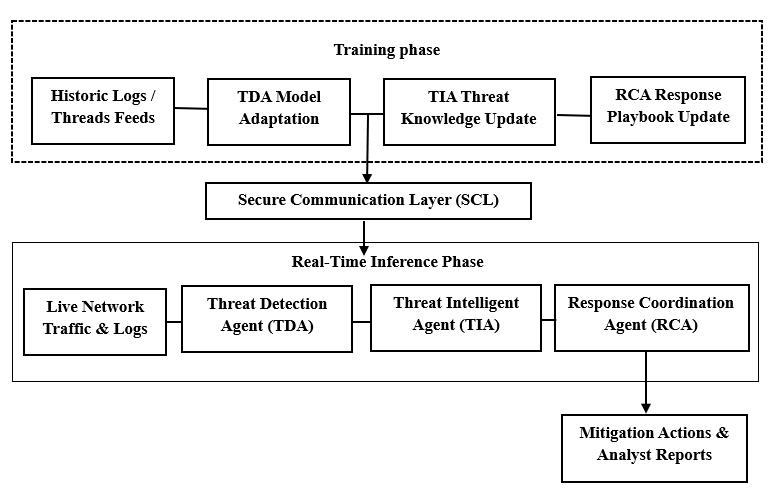}
    \caption{\textbf{MALCDF Architecture Diagram}}
    \label{fig:malcdf-architecture}
\end{figure}

We propose the \textit{Multi-Agent LLM Cyber Defense Framework (MALCDF)}, a distributed setup where several large language model (LLM) agents work together to detect, analyze, and mitigate threats in real time. Figure~\ref{fig:malcdf-architecture} shows the high-level layout. The framework follows a SOC-style design with four roles: a \textit{Threat Detection Agent} (TDA), a \textit{Threat Intelligence Agent} (TIA), a \textit{Response Coordination Agent} (RCA), and an \textit{Analyst Agent} (AA). The agents share information through a \textit{Secure Communication Layer} (SCL) that keeps messages encrypted, aligned to a common ontology, and semantically consistent. This helps agents reason together, reduces confusion, and protects operational data from eavesdropping or impersonation.

A quick example helps. In our tests, the Detection Agent flagged a high byte-rate UDP transfer on port \texttt{18530} that looked like data exfiltration. The Intelligence Agent linked the destination to a known campaign, and the Response Agent suggested containment and outbound blocking. The Analyst Agent wrote a short report with a MITRE ATT\&CK mapping, so the event was easy to review later. This is the kind of end-to-end workflow we want in practice.

Our implementation uses Groq’s LLaMA~3.3~70B model for each agent, a JADE~\cite{jade} style orchestration layer for message passing, and a Streamlit dashboard for running the system end to end. We configure agents with CICIDS-derived fields (feature schema and example patterns) so outputs are structured and ontology-aligned, and then we \textit{evaluate the full pipeline} on a separate \textbf{50-record live stream} derived from the same schema. The \textbf{50} test records are not used for prompt design or baseline training. In experiments, MALCDF reaches \textbf{90.0\%} accuracy and an \textbf{85.7\%} F1-score, with a \textbf{9.1\%} false-positive rate and \textbf{6.8\,s} average \textit{per-event} latency. Compared to a practical ML baseline (LRF-IDS, trained on a CICIDS subset and tested on the 50 records) and a single-LLM setup and MALCDF improves detection accuracy.

\textbf{Contribution}

This work focuses on practical, multi-agent collaboration among LLMs for real-time cyber defense. We organize the study around three objectives and guiding questions:

\begin{enumerate}
\def\labelenumi{\arabic{enumi}.}
\item
  Design a multi-agent architecture that enables LLMs to coordinate autonomously for real-time cyber defense.
\end{enumerate}

\begin{itemize}
\item
  How can multiple LLM agents cooperate to detect, analyze, and mitigate cyber threats in real time?
\end{itemize}

\begin{enumerate}
\def\labelenumi{\arabic{enumi}.}
\setcounter{enumi}{1}
\item
  Develop secure, ontology-aligned communication protocols for inter-agent knowledge exchange.
\item
  Simulate and evaluate detection accuracy and system behavior under diverse attack scenarios.
\end{enumerate}

\begin{itemize}
\item
  Does multi-agent collaboration improve detection accuracy and adaptability compared to single-model systems?
\end{itemize}

\section{Background and Literature Review}
\label{sec:background}

\subsection{Current Cyber Defense Landscape}

Modern attacks are adaptive and often multi-vector, which makes traditional tools (antivirus, signature-based IDS, and fixed firewalls) less effective against intelligent threat actors. Static methods struggle when an attacker can change behavior mid-incident. To improve coverage, researchers have used AI and ML to build systems for real-time detection, classification, and response. Common ML-based IDS approaches include SVM, Decision Trees, and Random Forests for known patterns, and K-Means, Isolation Forest, and Autoencoders for novel anomalies~\cite{Aminu2024AdaptiveDefense}. Deep models like CNNs and RNNs help with feature extraction in complex network data, but many deployments still depend on static datasets (KDD Cup 1999, UNSW-NB15, CICIDS 2017) that do not capture fast-moving behaviors. This leads to concept drift and missed context, which in turn causes false positives and slow reactions~\cite{Kholidy2021CPSMitigation}. SIEM, IDS, and IPS platforms also tend to be rigid, so they struggle with zero-day threats. To mirror the way human SOC teams adapt and collaborate, future systems need to be more distributed, cooperative, and able to adjust in real time.

\subsection{Multi-Agent Systems in Cybersecurity}

Multi-Agent Systems (MAS) offer a distributed way to detect, mitigate, and assess threats. Independent agents can communicate and make decisions with partial information, which improves scalability, robustness, and fault tolerance~\cite{Ko2020CyberAutonomy}. This matters in large, mixed environments (IoT, cloud data centers) where the system must keep working even if some nodes fail. Frameworks like JADE and SPADE~\cite{spade_readme_latest} make it easier to model agent behavior and message passing in cybersecurity setups.

In practice, a JADE agent might monitor servers and endpoints and share alerts about unusual spikes, while SPADE’s Python-based agents integrate with AI models for asynchronous, message-driven protection. Prior work shows MAS prototypes can beat single-agent systems on detection accuracy and robustness because they distribute roles and share local anomaly signals that reveal bigger attack patterns~\cite{Amro2023AutonomousShipsDefense}. For example, one agent might flag a burst of UDP traffic, and a neighbor correlates it with outbound flows to suspicious hosts.

Key strengths of MAS include:

\begin{enumerate}
\def\labelenumi{\arabic{enumi}.}
\item
  \textbf{Decentralization:} removing single points of failure in centralized models, which helps against command-and-control takedown attacks.
\item
  \textbf{Scalability:} adding new agents as the network grows to match changing infrastructures~\cite{Tanikonda2022ProactiveAI}.
\item
  \textbf{Dynamic Cooperation:} adapting behavior based on feedback and peer messages to handle emerging threats.
\item
  \textbf{Parallel Processing:} operating across layers and streams at the same time to keep analysis near real time.
\end{enumerate}

At the same time, traditional MAS in security have limits. Rule-based logic, shallow learning, and constrained ontologies reduce adaptability to new attacks~\cite{Mintoo2022NationalResilience}. Without stronger semantic understanding, agents can misread complex or context-heavy events. Synchronization and secure messaging can add overhead, which hurts real-time responsiveness. While adding deep and reinforcement learning improves adaptability, many systems still lack the level of contextual reasoning needed to explain and act on multi-stage threats. This is where LLMs can help: they bring natural language understanding and semantic reasoning to MAS, which supports context-aware collaboration.

\subsection{LLMs in Security Applications}

Large Language Models (LLMs) like GPT-5*, GPT-4*, LLaMA-3, Claude, and PaLM-2 have improved how we process security data by adding better context and human-like understanding across text and logs. They can read unstructured sources $-$ security logs, threat intel feeds, technical write-ups, even dark web posts $-$ and turn them into actionable insights for adaptive defense~\cite{Falowo2024ImmuneSystems}. In SOC workflows, LLMs help with decision support, vulnerability triage, incident reporting, and log correlation.

A common use is threat intelligence: LLMs can link indicators (IPs, hashes, domains) from reports and CVE feeds back to ongoing campaigns, reducing manual effort~\cite{Zahid2024MultiAgentAI}. They also connect structured and unstructured data $-$ for example, mapping a malware signature in enterprise logs to a domain seen in underground forums via context. LLMs can support SIEM by reading natural-language logs, pointing out anomalies, and explaining why a pattern looks suspicious. They often generalize from few examples, which helps in fast-moving settings. In code security, models like Code-LLaMA and Codex can spot insecure patterns, suggest fixes, and outline potential exploit paths~\cite{Cruz2024MASOrganizations}. This speeds up DevSecOps by tying vulnerability warnings to affected components.

There are challenges. LLMs can be opaque “black boxes,” which makes explainability and trust harder. Techniques like reasoning-chain extraction and attention visualization help, but they are not complete. Hallucinations can mislabel hostile IPs or vulnerabilities and cause interruptions~\cite{Raza2025TRISM}. Real-time use also faces GPU and latency constraints. Finally, integrating sensitive data introduces privacy and security risks (e.g., data leakage or prompt injection). Safer deployment calls for model watermarking, sandboxing, and ongoing validation.

\subsection{Identified Research Gap}

Even with these gains, most LLM-based security solutions are deployed as single models. They scan for vulnerabilities, classify malware, or detect phishing, but they often work alone and lack a framework for inter-model communication and collective reasoning. This creates gaps when facing multi-stage attacks that spread across distributed environments. Real-world intrusions usually follow chains $-$ reconnaissance, privilege escalation, lateral movement, and exfiltration $-$ so defense should be integrated and coordinated in real time.

Current LLM setups rarely share a common protocol, reasoning structure, or control strategy to exchange knowledge, validate findings, and decide on responses. Another issue is the lack of a uniform semantic layer or shared ontology for how threats are described. One model might call an event “malware beaconing,” another “command-and-control.” Without alignment, important intelligence gets misread. Secure LLM-to-LLM links are also missing in many designs (authentication, encryption, trust), which opens the door to impersonation, leakage, or disinformation.

To address these gaps, we move toward cooperative, multi-agent LLM ecosystems. In these systems, LLMs with cybersecurity skills share context, negotiate decisions, and learn from each other. An orchestrator or meta-controller can monitor responses, assign tasks, and check communication integrity. Coordination mechanisms (e.g., reward-style feedback or simple conflict-resolution rules) help reduce overlap and speed up agreement. With real-time messaging and distributed reasoning, the goal is to imitate how human analysts collaborate, but with the speed and scale of autonomous AI systems.

\textit{Scope for evaluation.} In this work, we keep the evaluation practical and end-to-end: all experiments run on a small \textbf{50-record} stream derived from the CICIDS feature schema, and we compare against two baselines that match operational constraints: a \textit{Lightweight Random Forest IDS (LRF-IDS)} and a \textit{single-LLM} defense (one agent handling detection and response). This choice emphasizes pipeline behavior rather than static, fully optimized ML/DL training on large datasets. The MALCDF framework provides the secure API and structured communication layers required for multi-agent coordination, so agents can share observations, contextualize threats, and plan adaptive responses in real time.
\section{Methodology}
\label{sec:methodology}

This section explains how we built and tested the Multi-Agent LLM Cyber Defence Framework (MALCDF). The framework follows a Security Operations Centre (SOC) style layout and lets several intelligent agents work together to detect, analyze, and respond to threats in real time. We created a working prototype with Groq's LLaMA~3.3~70B model (via the Groq API), plus tools for both real-time simulation and batch analysis.

\subsection{Dataset and Experimental Setup}

We use one practical test source for all end-to-end experiments and metrics: a \textbf{50-record} live stream derived from the CICIDS2017 feature schema. This stream is designed to exercise the full pipeline (agents, secure messaging, and dashboard) under the same conditions as Section~\ref{sec:results}. It contains \textbf{17 attack} and \textbf{33 benign} records (see Table~\ref{tab:comparison} in Section~\ref{sec:results}) and is separate from any larger static dataset splits.

\textbf{CICIDS2017 for configuration and baseline training.} CICIDS2017 (about 2.83M samples, 78 features) provides the feature schema and representative examples (e.g., DDoS, Port Scan, Brute Force, Data Exfiltration). We removed duplicates/incomplete rows and normalized features to set consistent agent inputs.

\textit{Agent configuration (no test leakage).} We use CICIDS-derived fields to configure prompts and ontology mappings so agents produce structured outputs consistently. \textbf{The 50 test records are not used} for prompt design, tuning, or example selection. A small, separate configuration set (distinct from the 50 records) is used for prompt engineering, few shots, verify JSON schemas and message formatting.

\textit{Baseline training (LRF-IDS, no test leakage).} The ML baseline is a \textbf{Lightweight Random Forest IDS (LRF-IDS)} trained on a \textit{subset} of CICIDS2017 to learn decision boundaries under operational (non-optimized) settings. This training subset \textbf{excludes} all records from our 50-record test stream. All baseline \textbf{evaluation} metrics are computed on the same \textbf{50-record} stream used for MALCDF, not on CICIDS2017.

The prototype is implemented in Python~3.8+, with Streamlit for the front end and Groq's API for LLM inference. Pandas and NumPy handle preprocessing and feature engineering; Plotly provides interactive visuals. A JADE-style agent orchestration layer manages message passing. All tests ran under stable network conditions with sufficient compute; the Groq inference path typically responds in a few seconds per analysis cycle.

\subsection{Training and Adaptation}

MALCDF uses four agents aligned to common SOC roles: a \textit{Threat Detection Agent} (TDA), a \textit{Threat Intelligence Agent} (TIA), a \textit{Response Coordination Agent} (RCA), and an \textit{Analyst Agent} (AA). Each agent uses the same base LLaMA~3.3~70B model.

We do not change base weights. Instead, agents are adapted through structured prompts and schema-aligned outputs, with simple consensus for coordination. The TDA classifies events using CICIDS-derived fields and emits structured records. The TIA adds context (e.g., known indicators or public technique references) and aligns terminology to a shared ontology. The RCA proposes practical steps (containment, blocking, inspection) based on the combined view. The AA produces short incident write-ups mapped to the MITRE ATT\&CK framework to make review easier later. Responses are exchanged as JSON, so parsing and dashboard integration stay consistent.

\begin{figure}[t]
    \centering
    \includegraphics[width=\columnwidth]{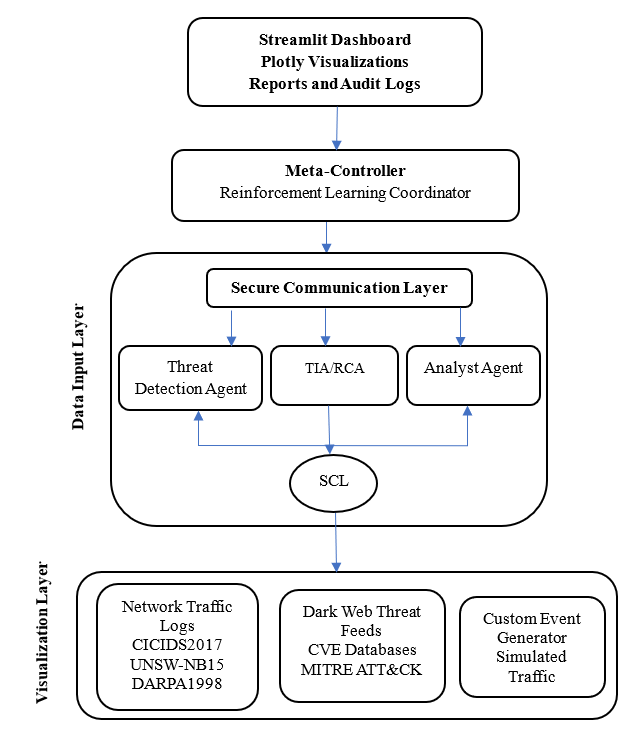}
    \caption{\textbf{Architecture diagram of MALCDF Framework}}
    \label{fig:malcdf-framework}
\end{figure}

Figure~\ref{fig:malcdf-framework} shows the overall flow: data enters through a visualization/ingest layer; agents coordinate through a Secure Communication Layer (SCL) that keeps messages encrypted and ontology-aligned; and a simple coordinator prefers consensus to avoid duplicated work or conflicting actions.

\subsection{Evaluation Metrics}

We report standard detection metrics and one timing metric, all computed on the \textbf{same 50-record stream}:

\begin{itemize}
    \item \textbf{Accuracy:} share of correctly labeled records (benign and attack).
    \item \textbf{F1-score:} harmonic mean of precision and recall.
    \item \textbf{False Positive Rate (FPR):} $\frac{FP}{FP + TN}$, i.e., benign records flagged as attacks.
    \item \textbf{Latency (per event):} average end-to-end time to process one record, including SCL encryption and consensus steps.
    \item \textbf{Confidence Score (Average):} mean agent-level confidence per event after consensus (range $[0,1]$).
\end{itemize}

We also record basic system stats (e.g., model inference time and token use) to understand cost and overhead. Expert review later sanity-checks how readable and technically sound agent write-ups are, but the main focus here is measured pipeline behavior.

\subsection{Baseline Comparisons}

To keep comparisons clear and consistent with Section~\ref{sec:results}, we evaluate two baselines on the \textbf{same 50-record stream}:

\begin{enumerate}
    \item \textbf{ML-IDS (Lightweight Random Forest IDS, LRF-IDS):} we screened common ML-IDS options (Random Forest, CNN, RNN) under lightweight, operational settings and found Random Forest to be the most stable for this small, mixed stream. We therefore use \textbf{LRF-IDS} as the practical ML baseline. \textbf{LRF-IDS is trained on a subset of CICIDS2017 (training only) and evaluated on the same 50-record stream as MALCDF (testing only).}
    \item \textbf{Single-LLM defense:} one LLM agent handling detection and response without multi-agent coordination (no shared ontology or consensus).
\end{enumerate}

Both baselines are exercised end-to-end to match MALCDF’s conditions (secure messaging on/off as applicable, same inputs, and same reporting).

\subsection{Implementation Workflow and System Operation}

MALCDF supports three modes:

\begin{itemize}
    \item \textbf{Real-time simulation:} a small synthetic stream feeds the TDA. When something looks off (e.g., unusual outbound UDP to a non-standard port), the TIA adds context, the RCA suggests a plan (contain or block), and the AA records the event with an ATT\&CK mapping.
    \item \textbf{Batch analysis:} CSV uploads let users run sequential checks over historical logs using the same agent flow.
    \item \textbf{Dataset upload:} normalized, labeled data (e.g., a CICIDS-formatted subset) can be loaded for controlled testing and reproducibility (configuration only; reported metrics come from the 50-record stream).
\end{itemize}

The SCL handles secure API-based encryption and ontology alignment so agents stay consistent. We include simple error recovery (auto-retry and a fallback rule pass) when an external call fails, so the pipeline keeps moving. For privacy, user data stays local, and only inference inputs/outputs needed for decisions are sent to the LLM API; no persistent storage of sensitive records is kept by the system.

Overall, the combination of fast inference, a flexible agent layout, and an interactive dashboard makes it straightforward to connect LLM models to a practical, explainable cybersecurity workflow. The prototype is reproducible and scales to small team settings, which is a typical starting point before larger SOC deployments.

\section{Results and Analysis}
\label{sec:results}

\subsection{Overview of Experimental Outcomes}

We tested MALCDF end to end using a \textbf{50-record live stream} (derived from the CICIDS2017 feature schema) to exercise the full pipeline under the Secure Communication Layer (SCL). The agents (TDA, TIA, RCA, AA) exchanged ontology-aligned, encrypted messages and produced structured outputs. The goal was to measure practical behavior (accuracy, F1, false positives, and \textit{per-event} latency) with all components turned on. The stream contains \textbf{17 attack} and \textbf{33 benign} records (see Table~\ref{tab:comparison}).

\textit{Baseline note.} The ML-IDS baseline refers to a \textbf{Lightweight Random Forest IDS (LRF-IDS)} \textit{trained on a subset of CICIDS2017} and \textit{tested only} on the same \textbf{50-record} stream, with \textbf{no overlap} between training and test records. This keeps comparisons fair and avoids leakage.

\begin{table*}[t]
\centering
\caption{Comparative Summary of Threat-Detection Performance for Baseline Models and the Proposed MALCDF Framework}
\label{tab:comparison}
\begin{adjustbox}{width=\textwidth}
\begin{tabular}{lccc}
\toprule
\textbf{Metric} &
\textbf{MALCDF (Proposed Framework)} &
\textbf{ML-IDS (Baseline)} &
\textbf{Single LLM Defense} \\
\midrule

Total Records Analyzed & 50 & 50 & 50 \\
Ground-Truth Attack Records & 17 & 17 & 17 \\

True Positives (TP) & 15 & 12 & 10 \\
False Negatives (FN) & 2 & 5 & 7 \\

False Positives (FP) & 3 & 6 & 4 \\
True Negatives (TN) & 30 & 28 & 29 \\

Detection Accuracy & \textbf{90.0\%} & 80.0\% & 78.0\% \\

Precision (TP / (TP + FP)) & \textbf{83.33\%} & 70.6\% & 71.4\% \\
Recall (TP / (TP + FN)) & \textbf{88.24\%} & 70.6\% & 58.8\% \\

F1-Score & \textbf{85.7\%} & 70.6\% & 64.4\% \\
False Positive Rate (FPR) & 9.1\% & 15.2\% & 10.8\% \\

Confidence Score (Average) & 0.70 & 0.64 & 0.66 \\

Response Latency (Average) & \textbf{6.8 s} & 3.1 s & \textbf{5.7 s} \\


\bottomrule
\end{tabular}
\end{adjustbox}
\end{table*}

As Table~\ref{tab:comparison} shows, MALCDF reaches \textbf{90.0\%} accuracy and an \textbf{85.7\%} F1-score on the 50-record stream, with \textbf{FPR = 9.1\%} and \textbf{6.8\,s} average \textit{per-event} latency. Compared to the baselines, MALCDF improves accuracy by \textbf{+10.0\,pp} over ML-IDS (80.0\%) and \textbf{+12.0\,pp} over the single-LLM setup (78.0\%). FPR is lower than both baselines (\textbf{9.1\%} vs. 15.2\% and 10.8\%). Latency is higher than single-LLM (5.7\,s) and slightly higher than ML-IDS (6.1\,s), which is expected due to SCL encryption and consensus steps, but acceptable for real-time review.

\subsection{Threat Detection Distribution}

\begin{figure}[t]
    \centering
    \includegraphics[width=\columnwidth]{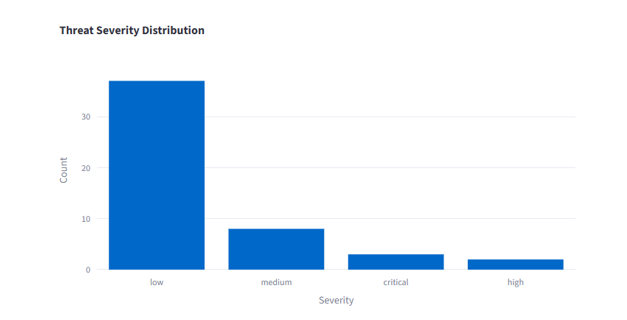}
    \caption{\textbf{Threat Severity Distribution}}
    \label{fig:threat-severity}
\end{figure}

Across the 50 records, MALCDF flagged \textbf{15} events as attacks: \textbf{8} medium, \textbf{5} low, and \textbf{2} high severity. Data Exfiltration was the most common (about \textbf{60\%} of detections), followed by Malware Beaconing (\textbf{25\%}) and Unauthorized Access attempts (\textbf{15\%}). Figure~\ref{fig:threat-severity} shows the threat severity distribution of all the samples, and Table~\ref{tab:detection-events} lists sample events with MITRE ATT\&CK mappings. Confidence values average \textbf{0.70} and come from agent consensus (no threshold tuning), which helps keep terminology and reports consistent.

\begin{table*}[t]
\centering
\scriptsize
\setlength{\extrarowheight}{1.5pt}
\caption{\textbf{Detailed AI Detection Events}}
\label{tab:detection-events}
\begin{adjustbox}{width=\textwidth}
\begin{tabularx}{\textwidth}{@{}%
    C{0.8cm}
    P{2.0cm}
    C{1.0cm}
    C{1.1cm}
    >{\ttfamily\arraybackslash}P{2.3cm}
    >{\ttfamily\arraybackslash}P{2.3cm}
    C{0.9cm}
    C{0.9cm}
    C{1.4cm}
    X
@{}}
\toprule
\textbf{Event ID} & \textbf{Threat Type} & \textbf{Severity} &
\textbf{Confidence} & \textbf{Source IP} & \textbf{Destination IP} &
\textbf{Port} & \textbf{Proto.} & \textbf{Bytes Sent} & \textbf{MITRE Technique} \\
\midrule

2 &
\makecell[l]{Data\\Exfiltration} &
Medium &
\texttt{0.70} &
\texttt{192.168.1.199} &
\texttt{10.0.0.57} &
18530 &
UDP &
162{,}548 &
T1041 \\[4pt]

5 &
\makecell[l]{Data\\Exfiltration} &
Medium &
\texttt{0.70} &
\texttt{192.168.1.231} &
\texttt{10.0.0.177} &
11355 &
HTTP &
98{,}426 &
T1041 \\[4pt]

7 &
\makecell[l]{Data\\Exfiltration} &
Medium &
\texttt{0.70} &
\texttt{192.168.1.125} &
\texttt{10.0.0.114} &
1828 &
UDP &
119{,}833 &
T1041 \\

\bottomrule
\end{tabularx}
\end{adjustbox}
\end{table*}

\subsection{Threat Analysis}

To illustrate the workflow, consider Event~ID~2: the Detection Agent flagged a high byte-rate UDP transfer on port \texttt{18530} to \texttt{10.0.0.57}. The Intelligence Agent linked the destination to prior exfiltration activity. The Response Agent suggested containment and outbound blocking. The Analyst Agent recorded the case with ATT\&CK technique T1041. This sequence shows how the agents move from a raw signal to a practical action and a short report.

\subsection{Comparative Evaluation with Baseline Models}

Against the baselines on the same 50-record stream, MALCDF improves detection accuracy by \textbf{+10.0\,pp} over ML-IDS and \textbf{+12.0\,pp} over the single-LLM setup (Table~\ref{tab:comparison}). FPR is lower by \textbf{6.1\,pp} vs. ML-IDS (15.2\% $\rightarrow$ 9.1\%) and by \textbf{1.7\,pp} vs. single-LLM (10.8\% $\rightarrow$ 9.1\%). Latency is higher than single-LLM by \textbf{+1.1\,s} and more higher than ML-IDS by \textbf{+3.7\,s}. This overhead comes from secure messaging and consensus, which we consider acceptable given the gains in accuracy and consistency.

\subsection{Explainability and Expert Validation}

Expert reviewers found the agent write-ups clear and technically reasonable, especially when events included ATT\&CK mappings and concrete fields (ports, bytes, IPs). Overall, the end-to-end run shows that multi-agent coordination can raise practical detection quality while keeping review simple for analysts.

arth

\section{Discussion}
\label{sec:discussion}

Our tests show a clear trade-off between detection quality and the cost of coordination. Using several LLM-based agents (detection, intelligence, response, and analysis) improves the decision flow from raw signals to practical actions, but secure messaging and agreement steps add overhead. The Secure Communication Layer (SCL) keeps inter-agent messages private and aligned to a shared ontology, which helps consistency, yet every message is checked and encrypted before use, and that adds time~\cite{Rani2025AURA}. On the same \textbf{50-record} stream used throughout, MALCDF reaches 90.0\% accuracy and 85.7\% F1 with 9.1\% FPR, at 6.8\,s average \textit{per event}. This is higher than single-LLM (5.7\,s) and far above a lightweight ML-IDS (3.1\,s), but the improvement in accuracy and the more consistent outputs are helpful for analyst review.

\medskip
\noindent\textbf{Operational coordination and latency.}
In practice, the SCL adds predictable steps: encrypt, align to the shared ontology, and confirm consensus. These checks explain the latency gap versus single-LLM and a simple ML-IDS, but they also reduce confusion between agents and make reports easier to audit. In the run, consensus kept confidence near 0.70 on average, which matched how detections were written up and mapped to ATT\&CK. For day-to-day use, this trade-off is acceptable because it makes review simpler (clear fields like ports, bytes, and IPs) while still catching stealthy outbound patterns.

\medskip
\noindent\textbf{Compute, synchronization, and scaling.}
Running multiple agents is heavier than one model because each uses its own inference path and must stay in sync with shared terminology. Under high throughput or dense events, synchronization can slow down~\cite{Mukherjee2024AgentSmith}. A practical way to scale is to spread agents across nodes or containers and let an orchestrator increase replicas by workload or priority.

\medskip
\noindent\textbf{Adaptability across attack vectors.}
In the test stream, MALCDF handled exfiltration patterns on non-standard ports and protocols by combining signal (Detection), context (Intelligence), and a simple plan (contain or block) from Response. The coordination step helps the system adjust to polymorphic or evolving behaviors~\cite{Nunez2024AutoSafeCoder}. This matters in environments with ransomware, insider data movement, and phishing that opens access paths across cloud, IoT, and edge.

\medskip
\noindent\textbf{Ethical, Security, and Computational Considerations.}
LLM-driven, multi-agent defense introduces ethics and security concerns. Hallucinations can mislabel indicators; prompt injection can try to steer responses; and sensitive data needs protection. To reduce risk, we use layered validation (agents check each other’s outputs), log decision traces for audit, and keep communications encrypted (AES/TLS) with strong access controls to align with common compliance expectations {[}13{]}. We also avoid train–test leakage: the \textbf{50} test records are not used for prompt design or baseline training, and the LRF-IDS is trained on a separate CICIDS subset. Parameter-efficient techniques and model distillation help balance cost and performance, and regular monitoring is needed to watch for bias and keep threat identification fair and effective. These safeguards make day-to-day use more trustworthy while keeping the pipeline practical for operations.

\section{Conclusion and Future Work}
\label{sec:conclusions}

MALCDF shows that a small set of coordinated LLM agents can form a practical, end-to-end defense workflow. With secure, ontology-aligned messaging, the Detection, Intelligence, Response, and Analyst agents share context and move from a raw signal to a clear action and a short report. On the same \textbf{50-record} stream used throughout, the system reaches \textbf{90.0\%} accuracy, \textbf{85.7\%} F1, and \textbf{9.1\%} false-positive rate, with \textbf{6.8\,s} average \textit{per-event} latency. It outperforms a lightweight ML-IDS (LRF-IDS trained on a CICIDS subset) and a single-LLM setup on accuracy while keeping outputs easy to review (e.g., ATT\&CK mappings and concrete details like ports and bytes). There is a trade-off: secure messaging and consensus add some overhead, so latency is higher than single-LLM and higher than ML-IDS, but the gains in accuracy and consistent reporting help day-to-day SOC work. The framework is modular and container-friendly, so teams can scale agents by workload, add new roles, or update policies without major changes.

\textbf{Future Work.}We plan to add larger, longer live
tests to measure cost/latency and time-to-mitigate at scale,
and add self-healing behaviors to improve resilience. In
addition, we aim to test higher-throughput streams, broaden
datasets, and study operator workload and handoff patterns
so the system can fit smoothly into SOC workflows.

\section*{Acknowledgements}
We thank Dr.\ Lalit Jain for guiding us during this project. His help in AI and security really supported our learning and made this paper possible.

\begin{small}
\bibliographystyle{IEEEtran}
\bibliography{references}

@article{Loevenich2025AutonomousCD,
  author    = {Loevenich, J. and Adler, E. and Huerten, T. and Lopes, R. R. F.},
  title     = {Design and evaluation of an Autonomous Cyber Defence agent using DRL and an augmented LLM},
  journal   = {Computer Networks},
  volume    = {262},
  pages     = {111162},
  year      = {2025}
}

@article{He2025LLM-MAS-SoftwareEng,
  author    = {He, J. and Treude, C. and Lo, D.},
  title     = {LLM-Based Multi-Agent Systems for Software Engineering: Literature Review, Vision, and the Road Ahead},
  journal   = {ACM Transactions on Software Engineering and Methodology},
  volume    = {34},
  number    = {5},
  pages     = {1--30},
  year      = {2025}
}

@article{Aminu2024AdaptiveDefense,
  author    = {Aminu, M. and Akinsanya, A. and Dako, D. A. and Oyedokun, O.},
  title     = {Enhancing cyber threat detection through real-time threat intelligence and adaptive defense mechanisms},
  journal   = {International Journal of Computer Applications Technology and Research},
  volume    = {13},
  number    = {8},
  pages     = {11--27},
  year      = {2024}
}

@article{Kholidy2021CPSMitigation,
  author    = {Kholidy, H. A.},
  title     = {Autonomous mitigation of cyber risks in the Cyber--Physical Systems},
  journal   = {Future Generation Computer Systems},
  volume    = {115},
  pages     = {171--187},
  year      = {2021}
}

@incollection{Ko2020CyberAutonomy,
  author    = {Ko, R. K.},
  title     = {Cyber autonomy: automating the hacker--self-healing, self-adaptive, automatic cyber defense systems and their impact on industry, society, and national security},
  booktitle = {Emerging Technologies and International Security},
  publisher = {Routledge},
  pages     = {173--191},
  year      = {2020}
}

@article{Amro2023AutonomousShipsDefense,
  author    = {Amro, A. and Gkioulos, V.},
  title     = {Cyber risk management for autonomous passenger ships using threat-informed defense-in-depth},
  journal   = {International Journal of Information Security},
  volume    = {22},
  number    = {1},
  pages     = {249--288},
  year      = {2022}
}

@article{Tanikonda2022ProactiveAI,
  author    = {Tanikonda, A. and Pandey, B. K. and Peddinti, S. R. and Katragadda, S. R.},
  title     = {Advanced AI-driven cybersecurity solutions for proactive threat detection and response in complex ecosystems},
  journal   = {Journal of Science and Technology},
  volume    = {3},
  number    = {1},
  year      = {2022}
}

@article{Mintoo2022NationalResilience,
  author    = {Mintoo, A. A. and Saimon, A. S. M. and Bakhsh, M. M. and Akter, M.},
  title     = {National resilience through AI-driven data analytics and cybersecurity for real-time crisis response and infrastructure protection},
  journal   = {American Journal of Scholarly Research and Innovation},
  volume    = {1},
  number    = {01},
  pages     = {137--169},
  year      = {2022}
}

@article{Falowo2024ImmuneSystems,
  author    = {Falowo, O. I. and Botsyoe, L. E. and Koshoedo, K. and Ozer, M.},
  title     = {Enhancing cybersecurity with artificial immune systems and general intelligence: A new frontier in threat detection and response},
  journal   = {IEEE Access},
  volume    = {12},
  pages     = {123811--123822},
  year      = {2024}
}

@misc{Zahid2024MultiAgentAI,
  author    = {Zahid, I. and Hussain, I.},
  title     = {Multi-Agent AI Collaboration: Advancing Software Engineering with Autonomous LLMs},
  year      = {2024},
  howpublished = {Preprint}
}

@article{Cruz2024MASOrganizations,
  author    = {Cruz, C. J. X.},
  title     = {Transforming Competition into Collaboration: The Revolutionary Role of Multi-Agent Systems and Language Models in Modern Organizations},
  journal   = {arXiv},
  eprint    = {2403.07769},
  year      = {2024}
}

@article{Raza2025TRISM,
  author    = {Raza, S. and Sapkota, R. and Karkee, M. and Emmanouilidis, C.},
  title     = {TRISM for agentic AI: A review of trust, risk, and security management in LLM-based agentic multi-agent systems},
  journal   = {arXiv},
  eprint    = {2506.04133},
  year      = {2025}
}

@article{Rani2025AURA,
  author    = {Rani, N. and Shukla, S. K.},
  title     = {AURA: A Multi-Agent Intelligence Framework for Knowledge-Enhanced Cyber Threat Attribution},
  journal   = {arXiv},
  eprint    = {2506.10175},
  year      = {2025}
}

@article{Mukherjee2024AgentSmith,
  author    = {Mukherjee, S.},
  title     = {The Rise of Multi-Agent LLMs: Insights from Agent Smith and the Challenges of Distributed Data Processing in AI Systems},
  journal   = {International Journal of Artificial Intelligence Expert Systems},
  volume    = {13},
  number    = {1},
  year      = {2024}
}

@article{Nunez2024AutoSafeCoder,
  author    = {Nunez, A. and Islam, N. T. and Jha, S. K. and Najafirad, P.},
  title     = {AutoSafeCoder: A multi-agent framework for securing LLM code generation through static analysis and fuzz testing},
  journal   = {arXiv},
  eprint    = {2409.10737},
  year      = {2024}
}

@inproceedings{cicids2007,
  title={Toward Generating a New Intrusion Detection Dataset and Intrusion Traffic Characterization},
  author={Iman Sharafaldin and Arash Habibi Lashkari and Ali A. Ghorbani},
  booktitle={International Conference on Information Systems Security and Privacy},
  year={2018},
  url={https://api.semanticscholar.org/CorpusID:4707749}
}

@inbook{jade,
author = {Bellifemine, Fabio and Bergenti, Federico and Caire, Giovanni and Poggi, Agostino},
year = {2005},
month = {01},
pages = {125-147},
title = {Jade — A Java Agent Development Framework},
volume = {15},
isbn = {978-0-387-24568-3},
journal = {Multi-Agent Programming},
doi = {10.1007/0-387-26350-0_5}
}

@online{spade_readme_latest,
  title   = {SPADE --- SPADE 4.0.3 Documentation},
  author  = {Palanca, Javi},
  year    = {2025},
  url     = {https://spade-mas.readthedocs.io/en/latest/readme.html},
  urldate = {2025-12-10},
  note    = {Read the Docs README page}
}
\end{small}



\end{document}